\documentclass[preprint,nofootinbib,aps,superscriptaddress,eqsecnum]{revtex4-1}

\usepackage[pdftex]{graphicx}
\usepackage{mathrsfs}
\usepackage{amssymb,amsmath,subfigure}

\newcommand{\lapprox}{%
\mathrel{%
\setbox0=\hbox{$<$}
\raise0.6ex\copy0\kern-\wd0
\lower0.65ex\hbox{$\sim$}
}}
\newcommand{\gapprox}{%
\mathrel{%
\setbox0=\hbox{$>$}
\raise0.6ex\copy0\kern-\wd0
\lower0.65ex\hbox{$\sim$}
}}

\newcommand{\ba}{\begin{array}}
\newcommand{\ea}{\end{array}}
\newcommand{\bd}{\begin{displaymath}}
\newcommand{\ed}{\end{displaymath}}
\newcommand{\beq}{\begin{equation}}
\newcommand{\eeq}{\end{equation}}
\newcommand{\bea}{\begin{eqnarray}}
\newcommand{\eea}{\end{eqnarray}}

\newcommand{\ra}{\rightarrow}

\def\q2 {q^2}

\def\bt{\begin{table}}
\def\et{\end{table}}

\catcode`@=11 
\def \gsim{\mathrel{\mathpalette\@versim>}}
\def \lsim{\mathrel{\mathpalette\@versim<}}
\def \@versim#1#2{\lower0.4ex\vbox{\baselineskip\z@skip\lineskip\z@skip
     \lineskiplimit\z@\ialign{$\m@th#1\hfil##\hfil$%
     \crcr#2\crcr\sim\crcr}}}
\catcode`@=12 

\begin{document}

\renewcommand*{\thefootnote}{\fnsymbol{footnote}}

\begin{center}

{\large\bf  Obstacles to extending $R$-parity violation 
to Supersymmetric SU(5)}\\[15mm] 
Najimuddin Khan $^{a,}$\footnote{E-mail: phd11125102@iiti.ac.in}, Subhendu
Rakshit $^{a,}$\footnote{E-mail: rakshit@iiti.ac.in} and Amitava
Raychaudhuri $^{b,}$\footnote{E-mail: palitprof@gmail.com}
\\[2mm]

{\em $^a$Discipline of Physics, Indian Institute of Technology Indore,\\
 Khandwa Road, Simrol, Indore - 452~020, India}
\\[2mm] 
{\em $^b$Department of Physics, University of Calcutta,  
92 Acharya Prafulla Chandra Road, Kolkata 700~009, India}
\\[20mm]
\end{center}

\begin{abstract} 

\vskip 20pt

We explore the consequences of promoting bilinear $R$-parity
violation, usually formulated in the minimal supersymmetric
standard model framework, to a supersymmetric SU(5) grand unified
theory. We observe that  the limits on proton decay  and neutrino
mass  place  tight constraints on the bilinear SU(5) $R$-parity
violating parameters  creating a different doublet-triplet issue
which cannot be resolved by an extension of the usual fine-tuning
in the symmetry breaking scalar sector.  If the parameters 
are made to satisfy the constraints, albeit unnaturally, there
remains no room for the possibility to correct the SU(5) fermion
mass ratios by introducing $R$-parity violation.

\end{abstract}

\vskip 1 true cm
\maketitle

\setcounter{footnote}{0}
\renewcommand*{\thefootnote}{\arabic{footnote}}

\section{Introduction}
In the standard model (SM), the Yukawa  couplings are hand-picked
to explain the observed fermion masses via the Higgs mechanism. In
the minimal supersymmetric standard model (MSSM), fermion masses are
obtained from two Higgs fields $H_u$ -- which gives mass to the
up-type quarks -- and $H_d$ -- which is responsible for the
masses of the down-type quarks and charged leptons. Due to the absence
of the right-handed fields the neutrinos cannot acquire a Dirac
mass. Further, if lepton number conservation is imposed then this
forbids a neutrino Majorana mass. 

When this model is embedded in a grand unified theory (GUT), such
as SU(5), the unification of couplings is an attractive consequence.
The SU(5) symmetry also requires the ratio of down-type quark and
charged lepton masses  of each generation to be unity at the GUT
scale.
\beq
\left(\frac{m_d}{m_\ell}\right)_i= 1 \qquad i=1,2,3 \, ,
\eeq
while the ratios obtained by extrapolating the measured masses are
\beq
\frac{m_d}{m_e} \sim 2.6\, ,\quad \frac{m_s}{m_\mu} \sim 0.23\,
,\quad \frac{m_b}{m_\tau} \sim 0.81 \, .
\eeq

If the requirement of $R$-parity symmetry is relaxed one of the
main motivations of MSSM -- the LSP dark matter candidate -- is
lost. But  $R$-parity violation (RPV) also has its virtues. It
can be used to explain the observed pattern of neutrino masses
and mixings~\cite{RPV-neutrino, Rakshit:2004rj}. Hence it is
pertinent to ask the question whether RPV can address the
mismatch of wrong fermion mass ratios posed in 
supersymmetric (SUSY) SU(5)~\cite{Bajc:2015zja}. The issue can
be alleviated using the trilinear $A$
terms~\cite{DiazCruz:2000mn}.  Alternatively, a solution can be
obtained by adding  $5+\bar{5}$ vector-like matter fields in SUSY
SU(5)~\cite{Babu:2012pb}.  For other approaches using non-minimal
models see Ref.~\cite{others}.

Non-observation of proton decay, e.g.,  at SuperKamiokande
\cite{Nishino:2009aa, Abe:2014mwa}, poses severe
constraints~\cite{protondecay} on grand unified theories.  In
non-SUSY theories, typified by SU(5), proton decay is driven by
dim-6 operators. In $R$-parity conserving SUSY SU(5), the
existence of sfermions at the electroweak scale allows proton
decay to proceed through  dim-5 operators~\cite{Weinberg:1981wj,
Sakai:1981pk, Bajc:2002bv}.  However, when RPV is admitted proton
decay can  arise even from dim-4 operators, which puts severe
restrictions on the size of $R$-parity violating
parameters~\cite{Smirnov:1995ey, Bhattacharyya:1997vv,
Bhattacharyya:1996nj}.

Neutrino mass is another area where RPV interactions which
violate lepton number can play an important role. RPV results in
mixing between neutralinos and  neutrinos. This leads to one
neutrino state becoming massive~\cite{RPV-neutrino,
Rakshit:2004rj}. The observed smallness of neutrino masses limit
the size of $R$-parity violating interactions.

In this paper we show that extension of bilinear $R$-parity
violation of MSSM to the SU(5) theory faces a serious obstacle in
maintaining consistency with proton decay and neutrino mass
constraints.   Further it is {\em not} possible to find a
satisfactory resolution of the issue of wrong fermion mass ratios
within SUSY SU(5) even in the context of $R$-parity violation
unless severe accidental fine-tunings amongst various
uncorrelated sectors are entertained~\cite{Bajc:2015zja}.

\section{RPV SUSY SU(5): A Flashback}
In minimal SUSY SU(5) the matter fields -- all left-handed -- 
are contained in
\beq
\bar{5}_i \equiv \underbrace{(\bar{3},1)}_{d^c_{0i}} +
                  \underbrace{(1,2)}_{L_{0i}}
\qquad 
{\rm and}
\qquad
10_i \equiv \underbrace{(\bar{3},1)}_{u^c_{0i}} +
            \underbrace{(3,2)}_{Q_{0i}} +
            \underbrace{(1,1)}_{e^c_{0i}}   \;,
\eeq 
where $i=1,2,3$ is the generation index and the numbers in the parentheses are the SU(3)$_c$ and SU(2)$_L$ quantum numbers.  $L_1=(\nu_e,e)_L$ and
$Q_1=(u,d)_L$ stand for left superfields. $d^c\equiv (d_R)^C$,
where $d_R$ is the right chiral down quark superfield. The same
is true for the other right superfields $u_R$ and $e_R$. The
subscript $0$ is indicative of the flavour basis.
Colour indices are suppressed.  We express the above in the form:
\beq 
\bar{5}_i= \begin{pmatrix}
d^c_0 \\ \epsilon_2 L_0 
\end{pmatrix}_i 
\qquad
10_i=\begin{pmatrix} 
\epsilon_3 u^c_0 &  Q_0 \\ -Q^T_0 & \epsilon_2 e^c_0 
\end{pmatrix}_i \;.
\eeq 
$\epsilon_n$ represents the $n$-dimensional completely
antisymmetric tensor with $\epsilon_{12} = +1$ and
$\epsilon_{123} = +1$.

The Higgs fields are contained in
\beq
5_H= \begin{pmatrix}
T \\  H_u
\end{pmatrix}  
\qquad 
\bar{5}_H= \begin{pmatrix}
\overline{T} \\  H_d
\end{pmatrix} \, .
\eeq  
The scalar field which breaks SU(5) to the SM  resides
in a 24-plet adjoint representation. $T$,  $\overline{T}$
represent colour triplets having masses around the GUT scale,
$M_{GUT}\sim 10^{16}$~GeV. 

Once $R$-parity violation is considered, there is no quantum
number that distinguishes $\bar{5}_i$ from $\bar{5}_H$ and as a
result it is convenient to club these using the following
notation
\beq
\bar{5}_\alpha = \begin{pmatrix}
\bar{3}_\alpha \\  \bar{2}_\alpha
\end{pmatrix}   \qquad \alpha=0,1,2,3 \;,
\eeq
with $\bar{5}_0=\bar{5}_H$.
We can then write the superpotential keeping only the relevant
terms for this discussion as\footnote{The superpotential has the
matter fields in the flavour basis. We have suppressed the
subscript $0$ to avoid cluttering of the notation.}
\beq
W \in \bar{5}_\alpha \,(M_\alpha+\eta_\alpha 24)\, 5_H +
\frac{1}{2} Y^5_{\alpha \beta k}  \bar{5}_\alpha \bar{5}_\beta
10_k + Y^{10}_{ij} 10_i 10_j 5_H \; . \label{master} 
\eeq  
$Y^5_{\alpha \beta k}$ is antisymmetric in the first two indices
while $Y^{10}_{ij}$ is symmetric under $i\leftrightarrow j$. 
Yukawa couplings for $H_u$ and $H_d$ are obtained from
$Y^{10}_{ij}$ and $Y^5_{0jk}$ respectively. We choose the fermion
basis states so that the latter is diagonal, i.e., $Y^5_{0jk} =
Y^5_j \delta_{jk}$. $Y^5_{ijk}$ are
trilinear RPV couplings which we take to be absent, the entire
$R$-parity violation arising from the bilinear mixing encoded in the
first term in Eqn. (\ref{master}). We remark later about the
possibility of keeping $Y^5_{ijk}$ non-zero and fine-tuning them to  
cancel off the effects arising from the bilinear $R$-parity violation.

$M_\alpha$ represent SU(5)-invariant mass terms and their
natural scale is  
${\cal O}(M_{GUT})$. The coupling $\eta$ is
expected to be  ${\cal O}$(1) or smaller.

The mass terms of the superpotential
are given by
\beq
\bar{3}_\alpha {\cal M}_\alpha T + \bar{2}_\alpha \mu_\alpha H_u \, ,
\label{eq:mix}
\eeq
with
\bea
{\cal M}_\alpha &=& M_\alpha + 2 \eta_\alpha V \label{Mi} \; , \\
\mu_\alpha &=&  M_\alpha -3  \eta_\alpha V \label{mui}\; ,
\eea
where $V$ is the vacuum expectation value (VEV) received by the
24-plet scalar field around the GUT scale.  ${\cal M}_i$
($\mu_i$) stand for bilinear RPV couplings involving colour
triplets (SU(2) doublets). In SUSY with RPV neutrinos (sneutrinos) mix
with neutralinos (neutral Higgs) and charged leptons (sleptons)
mix with charginos (charged Higgs) due to the presence of
$\mu_i$. When RPV SUSY is embedded in an  SU(5) GUT the additional
bilinear RPV couplings ${\cal M}_i$ allow the fermionic
members of the colour triplet superfields
$T$ and $\overline{T}$ (scalars of $T$ and $\overline{T}$) to  mix
with down-type quarks (squarks).  We will mainly focus on the
phenomenology of these new RPV couplings in SUSY SU(5).

So long as SU(5) is exact, both ${\cal M}_\alpha$ and
$\mu_\alpha$ are forced to lie at the same scale as ${M}_\alpha$.
When SU(5) breaking occurs, the 24-plet acquires a GUT-scale
VEV. The challenge to keep ${\cal M}_0$ at the GUT level while
maintaining $\mu_0$ at the electroweak scale is known as the
doublet-triplet splitting problem and it calls for a large
fine-tuning of the $R$-parity conserving term. Specifically, $\eta_0$ can be
fine-tuned such that a cancellation occurs in Eqn.~(\ref{mui})
between the two ${\cal O}(M_{GUT})$ terms leaving a tiny $\mu_0
\sim {\cal O}(M_{W})$ while from Eqn.~(\ref{Mi}) ${\cal M}_0$
remains at the GUT scale. The same equations appear to leave
open the option to similarly fine-tune the bilinear RPV-terms
${\cal M}_i$ and $\mu_i$ through the $\eta_i$. However, it is
clear from the difference in sign in the two equations that if
one of ${\cal M}_i$ and $\mu_i$ is fine-tuned to a small value
the other must remain at the GUT scale.  In this note we stress
that {\em both} ${\cal M}_i$ and $\mu_i$ are required to be
significantly smaller than $M_{GUT}$ from the limits on proton
decay and from the neutrino mass scale, respectively, which
cannot be realised in the above manner.  Thus an extension of
bilinear $R$-parity violation to SU(5) is fraught with an
inherent hurdle.

The Lagrangian containing the mass terms for the colour triplet
fermions including the contribution from the first term in Eqn.
(\ref{eq:mix}) can be schematically written as\footnote{A typical
fermion mass term can be expressed as $m_{12}
\{\psi_{f_1}^C\}^{T}  {\cal C}^{-1}  \psi_{f_2}$  where ${\cal C}$
is the charge conjugation operator. For chiral fermions this is
$m_{12} \{\psi_{f_1}^C\}^{T} P_L {\cal C}^{-1} P_L \psi_{f_2}$ +
h.c.  which is written here in condensed form as $m_{12}
\{f_1^c\}_L \{f_2\}_L$.}:
\beq
\begin{pmatrix}
\overline{T}_0 & d^c_{0i} 
\end{pmatrix}
\begin{pmatrix}
{\cal M}_0 & 0 \\ 
{\cal M}_i & m^{\rm diag} 
\end{pmatrix} 
\begin{pmatrix}
T_0 \\ d_{0j}
\end{pmatrix} \; , \label{quarkT} 
\eeq
where,  due to the choice of the $d$-type quark basis,
$(m^{\rm diag})_{ij} =\delta_{ij}\,Y^5_{j}\,v_d $  is a $3\times
3$ diagonal matrix ($v_d$ being the vacuum expectation value of
$H_d$). The mass matrix in (\ref{quarkT}) can be diagonalised by
a bi-unitary transformation (see Appendix ~\ref{appendix1}):
\beq
\begin{pmatrix}
\overline{T}\\ d_i^c
\end{pmatrix}=
\begin{pmatrix}
U_{00} & U_{0j} \\ 
U_{i0} & U_{ij}\\ 
\end{pmatrix}
\begin{pmatrix}
\overline{T}_0\\ d_{0j}^{c} 
\end{pmatrix}
  \qquad {\rm and}   \qquad
\begin{pmatrix}
T\\ d_i 
\end{pmatrix}=
\begin{pmatrix}
V_{00} & V_{0j} \\ 
V_{i0} & V_{ij}
\end{pmatrix}
\begin{pmatrix}
T_0 \\ d_{0j}
\end{pmatrix}\;.
\label{mprime}
\eeq
In the mass basis Eqn.~(\ref{quarkT}) becomes:
\beq
\begin{pmatrix}
\overline{T}& d_i^{c}
\end{pmatrix}
\begin{pmatrix}
 M_T & 0 \\ 
0 & m_{d}\\ 
\end{pmatrix}
\begin{pmatrix}
T\\ d_i 
\end{pmatrix} \; ,
\label{massb}
\eeq
with $m_d=diag(m_{d_i})$, $m_{d_i}$ being the down-type quark
masses at the grand unification scale.

In Eqn.~(\ref{mprime}) $V$  is close to the identity matrix, as
illustrated in the one generation  case in
Appendix~\ref{appendix1}.

The matrix $U$ is then given by~\cite{Babu:2012pb, Bajc:2015zja} 
\bea
U_{00} &=& \frac{1}{\sqrt{1+|x_i|^2}}  \;, \label{Umatrix1}\\
U_{0i} &=& -U_{i0}=\frac{x_i}{\sqrt{1+|x_i|^2}} \;,  \label{Umatrix2}\\
U_{ij} &=& \delta_{ij}-\frac{x_i
x_j}{\sqrt{1+|x_i|^2}\,(1+\sqrt{1+|x_i|^2})} \;, \label{Umatrix3}
\eea
where $x_i={\cal M}_i/{\cal M}_0$ and $|x_i|^2 \equiv \sum_i x_i^2$.

So far we have focussed on the first term of Eqn.
(\ref{eq:mix}) which mixes SU(2) singlet, colour anti-triplets,
$\overline{T}$, and the $d_i^c$ states. The second term of the equation
produces the commonly considered bilinear $R$-parity violating
mixing between the $H_d$ and the $L_i$ which are colour singlet,
SU(2) doublets.  The upshot of this is mixing between the
charged leptons and the chargino on the one hand and the neutrinos
and the neutralino on the other.

For illustration, the chargino
mass matrix is extended to incorporate mixing of the charged
leptons with the superpartner of the Higgs $H_d$.  Denoting by
$M_2$ the SU(2) gaugino mass, the extended mass matrix
can be written as
\beq
\begin{pmatrix} \widetilde{W}^{-}_0 & \widetilde{H}_{0d}^- & e_{0i}
\end{pmatrix}
\begin{pmatrix}
M_2 & g v_u/\sqrt{2} & 0 \\
g v_d/\sqrt{2} & \mu_0 & 0\\
0 & \mu_i & m^{\rm diag}
\end{pmatrix}
\begin{pmatrix}
\widetilde{W}^+_0 \\ \widetilde{H}_{0u}^+ \\ e^c_{0j}
\end{pmatrix} \;.
\label{chargino}
\eeq
It is important to note at this point that the same $m^{\rm
diag}$  appears here as in Eqn.~(\ref{quarkT}) due to the SU(5)
symmetry which in turn predicts wrong fermion mass relations
$m_{d_i}=m_{\ell_i}$ in $R$-parity conserving SU(5) supersymmetry.
We will explore whether due to the presence of RPV bilinear
couplings ${\cal M}_i$ and $\mu_i$, the ratios correct
themselves.

Experimental evidences accumulating at the LHC exclude gluino
masses of a TeV or less. If $M_2$ is also greater than 1 TeV then the
wino-state essentially decouples in Eqn. (\ref{chargino}).
Accordingly, let us concentrate on the following submatrix in
Eqn.~(\ref{chargino}) which has a similar structure as the mass
matrix in Eqn.~(\ref{quarkT}):
\beq
\begin{pmatrix} \widetilde{H}_{0d}^- & e_{0i} \end{pmatrix}
\begin{pmatrix}
 \mu_0 & 0\\
 \mu_i & m^{\rm diag}
\end{pmatrix}
\begin{pmatrix}
\widetilde{H}_{0u}^+ \\ e^c_{0j}
\end{pmatrix}
\, . \label{subchargino}
\eeq
Here $\mu_0$ is at the electroweak scale -- the lighter
scale arising from the fine-tuning of $\eta_0$ for {\em doublet-triplet
splitting} alluded to earlier.  To start with we consider $\mu_i$
to be kept at the same order by a similar tuning of $\eta_i$, which
according to Eqn. (\ref{mui}) keeps
${\cal M}_i$ at the GUT scale\footnote{We show later that the
smallness of the neutrino mass calls for a far smaller $\mu_i$,
i.e., a higher degree of fine-tuning.}.

This matrix is diagonalised by going to the mass basis:
\beq
\begin{pmatrix}
\widetilde{H}_d^- \\ e_i
\end{pmatrix}=
\begin{pmatrix}
\widetilde{U}_{00} & \widetilde{U}_{0j} \\ 
\widetilde{U}_{i0} & \widetilde{U}_{ij}\\ 
\end{pmatrix}
\begin{pmatrix}
\widetilde{H}_{0d}^{-} \\ e_{0j}
\end{pmatrix}
  \qquad {\rm and}   \qquad
\begin{pmatrix}
\widetilde{H}_u^+ \\ e^c_i 
\end{pmatrix}=
\begin{pmatrix}
\widetilde{V}_{00} & \widetilde{V}_{0j} \\ 
\widetilde{V}_{i0} & \widetilde{V}_{ij}\\  
\end{pmatrix}
\begin{pmatrix}
\widetilde{H}_{0u}^{+} \\ e^{c}_{0j} 
\end{pmatrix}\;.
\label{mprimel}
\eeq
The matrix $\widetilde{U}$  is obtained by the replacement $x_i
\rightarrow y_i = \mu_i/\mu_0$ in Eqns. (\ref{Umatrix1}) - (\ref{Umatrix3}) and
$\widetilde{V} \simeq {\cal I}$ so long as $\mu_i \sim \mu_0 \gg m_{d_i}$.

The mixing in the neutralino - neutrino sector -- both
Majorana fields -- is well
studied and, as is well known, leads to one neutrino state
getting a non-zero mass. We return to this later.

The message from this analysis is that an extension of bilinear
$R$-parity violation to SU(5) leads to mixing between the colour
anti-triplets $\overline{T}$ and $d_i^c$ -- Eqn. (\ref{mprime}) --
besides the much studied $\widetilde{H}_d^- - e_i$ mixing given
in Eqn. (\ref{mprimel})  and a similar mixing in the neutrino -
neutralino sector.  The natural magnitude of these mixings is
${\mathcal O}$(1). As for the usual doublet-triplet mixing, it is
possible through fine-tuning to make one of these, but not both,
to be small.

\section{RPV SUSY SU(5): Trilinear couplings}

The Yukawa couplings for $\overline{T}$ arise from the $Y^5$ term in
Eqn.~(\ref{master}) which can be written as 
\beq
Y^5_{i} \left[\overline{T}_\rho
\{\epsilon_{\rho\xi\sigma} ~d_{i\xi}^c ~u_{i\sigma}^c 
+ ~(u_{i\rho}~e_i-d_{i\rho}~\nu_{e_i}) \} 
- H_d^- \{d_{i\alpha}^c u_{i\alpha} + ~\nu_{e_i} e_i^c\}
+ H_d^0\{d_{i\alpha}^c d_{i\alpha} + ~e_i e_i^c\}\right] \; ,
\eeq
where $\rho$, $\xi$, $\sigma$ are colour indices.

When expressed in the mass basis using Eqns.~(\ref{mprime}) and
~(\ref{mprimel}), it generates the $\lambda$, $\lambda^\prime$ and
$\lambda^{\prime\prime}$ trilinear couplings in the following RPV
superpotential
\beq
    {\cal W}_{\not R}  =  {1\over 2}\lambda_{ijk} L_i L_j e^c_k
                        +  \lambda'_{ijk} L_i Q_{j\rho}  d^c_{k\rho}    
                        +  {1\over 2}\lambda''_{ijk}\epsilon^{\rho\xi\sigma} u_{i\rho}^c   d_{j\xi}^{c} d_{k\sigma}^{c} 
                        + \mu_i L_i H_u,
\eeq
where 
\bea
\lambda_{iik}^{\prime}&=& - Y^5_{i}~U_{k0} \; , \label{lamprime}\\
\lambda_{ijk}^{\prime\prime}&=&
- Y^5_{i}~(U_{k0}~U_{ij}-U_{j0}~U_{ik}) \; . \label{lampprime}
\eea
Here we neglect rotation due to $V$ and, as noted earlier, work in a
basis where the matrix $Y^5_{0ij}$ is diagonal with the
elements  proportional to the down-type quark or
charged lepton masses. We note that not all the $\lambda^\prime$
couplings are obtained.  $\lambda$-type couplings are generated
due to the bilinear mixing terms $\mu_i$. These mixings also
modify  Eqns.~(\ref{lamprime} - \ref{lampprime}). As the $\mu_i$ turn
out to be rather small, we do not display these effects here.
RPV originally contained in ${\cal M}_i$ is now manifested in the
form of trilinear RPV couplings as above.

\section{Neutrino mass and Proton decay constraints}

As mentioned earlier, the presence of non-zero $\mu_i$ leads to
neutrino-neutralino mixing  and the neutralino mass matrix is
extended to incorporate the massless neutrinos. During
diagonalisation of this matrix the small $\mu_i$ terms along
with the much heavier neutralino masses lead to a seesaw-like
contribution to neutrino masses.  This way only one neutrino (can
be chosen as the heaviest one) becomes massive whereas the other
neutrinos can get loop level contributions in the RPV scenario.
A rough estimate yields $m_\nu \sim \sum_i
\mu_i^2/\tilde{m}$~\cite{RPV-neutrino, Rakshit:2004rj}, where
$\tilde{m}$ represents neutralino mass and is set to $\mu_0\sim
1$~TeV. To reproduce $m_\nu \sim 0.1$~eV,  one requires $\mu_i
\sim 10^{-3}$~GeV,  i.e. $y_i = (\mu_i/\mu_0) \sim 10^{-6}$. Due to
this hierarchy between $\mu_0$ and $\mu_i$, when the  charged
fermion mass matrix in (\ref{subchargino}) is diagonalised,
$m^{\rm diag} \sim diag({m_{\ell_i}})$ implying $Y^5_{i}\simeq
m_{\ell_i}/v_d$ in a basis in which $Y^5_{0ij}$ is diagonal.

\begin{figure}[h!]
 \begin{center}
\subfigure[]{
 \includegraphics[width=2.8in,height=2in, angle=0]{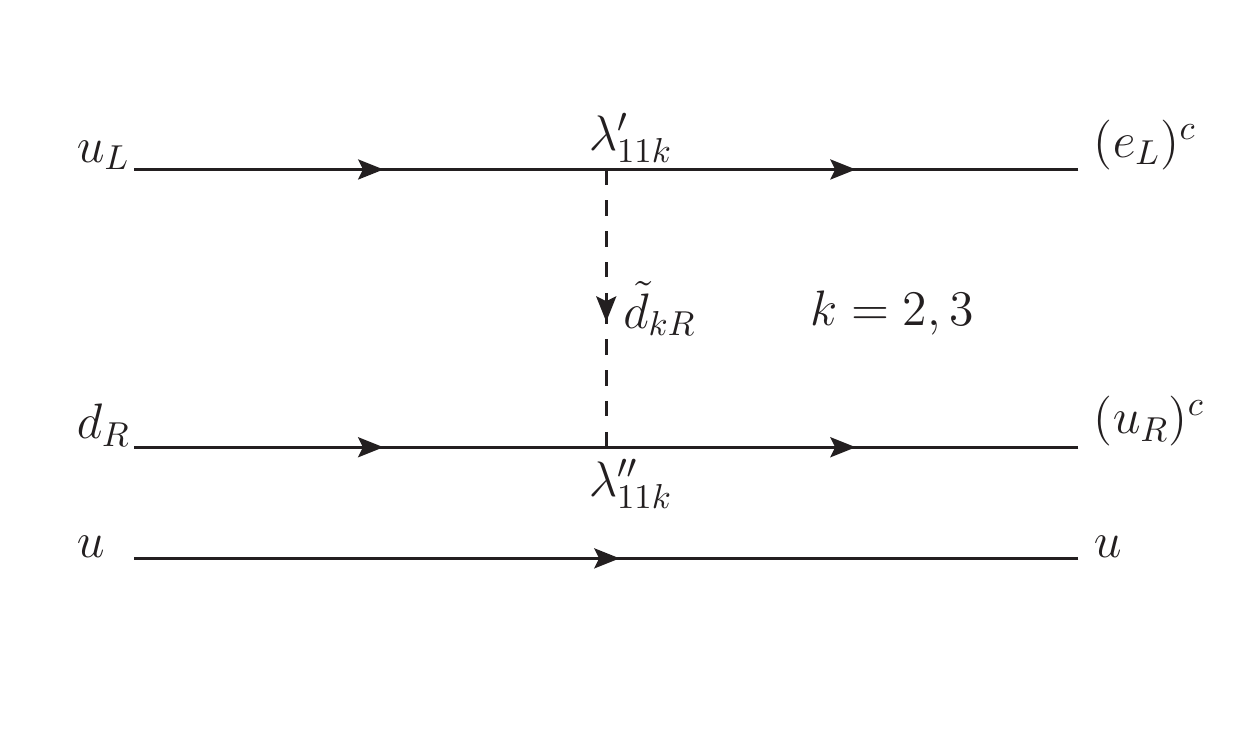}}
 \hskip 1pt
 \subfigure[]{
 \includegraphics[width=2.8in,height=2in, angle=0]{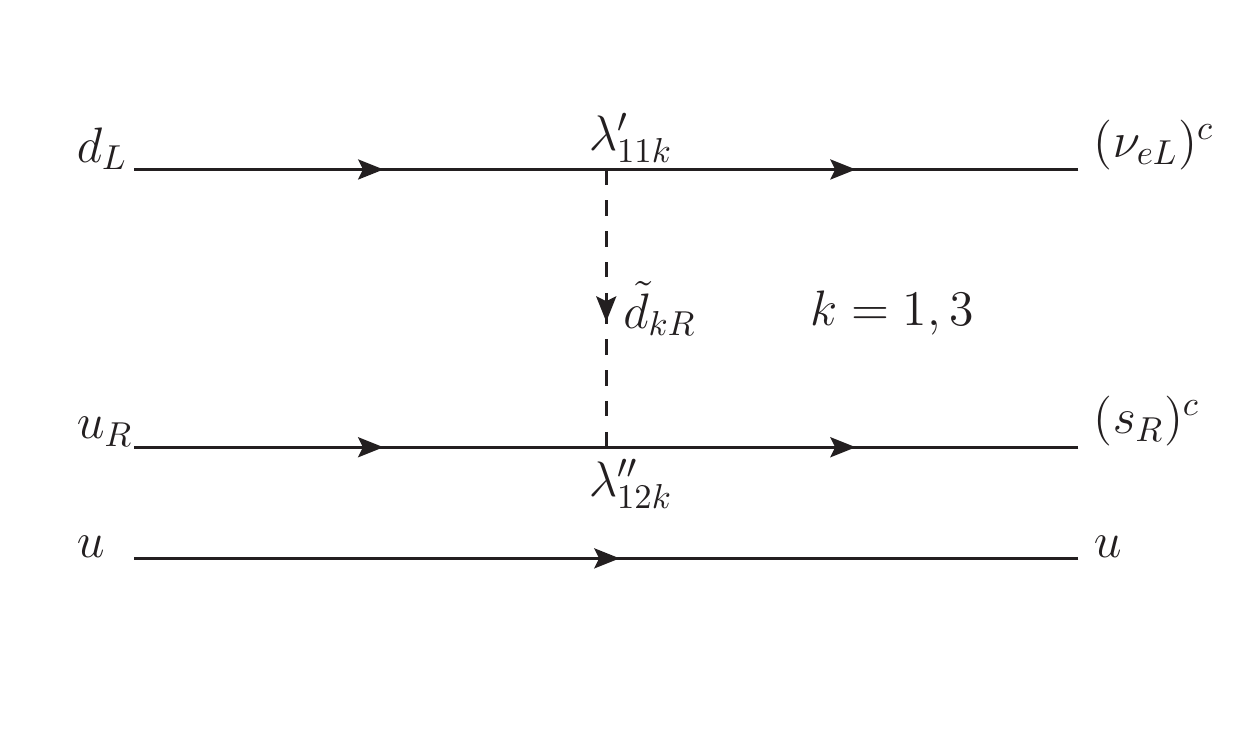}}
 \caption{\label{fig:Protondecay} \textit{Proton decay channels
(a) $p\ra e^+ + \pi^0$ and (b)  $p\ra \bar{\nu}_e + K^+$ mediated by
trilinear RPV couplings.}}
 \end{center}
 \end{figure}

In the presence of both $\lambda^\prime$ and $\lambda^{\prime\prime}$
couplings, proton decay can proceed through $p\ra e^+ + \pi^0$
and $p\ra \bar{\nu}_e + K^+$ as shown in Fig.~\ref{fig:Protondecay}.
The SuperKamiokande collaboration has put lower limits~\cite{Nishino:2009aa, Abe:2014mwa} on  the
proton lifetime for decays {\em via} $p\ra e^+ + \pi^0$ and $p\ra \nu_e +
K^+$  channels as $8.2 \times 10^{33}$ years and $5.9 \times
10^{33}$ years respectively. These limits  constrain the strength
of RPV \cite{Barbier:2004ez} and we find: 
\beq
|\lambda_{11k}^{\prime}~\lambda_{11k}^{\prime\prime}|<2.0 \times
10^{-25} \qquad {\rm and} \qquad
|\lambda_{11k}^{\prime}~\lambda_{12k}^{\prime\prime}|<1.6 \times
10^{-25} \;,
\eeq
for squark mass $\tilde{m} = 1$~TeV.

Using Eqn.~(\ref{lamprime}), the above constraints can be used to write
\beq
|(Y^5_{1})^2 ~ U_{k0}(U_{k0}~U_{11}-U_{10}~U_{1k})| < 2.0 \times
10^{-25} \; , \qquad k=2,3 \; ,\\
\label{bound11}
\eeq
\beq
|(Y^5_{1})^2 ~ U_{k0}(U_{20}~U_{1k}-U_{k0}~U_{12})| < 1.6 \times
10^{-25} \; , 
\qquad k=1,3 \; .
\label{bound12}
\eeq
Now using Eqn.~(\ref{Umatrix1} - \ref{Umatrix3}), one can restrict the bilinear RPV
terms involving $\overline{T}$ as $x_1<1.7 \times 10^{-7}$,
$x_2<2.1\times10^{-7}$ and $x_3<2.1\times10^{-7}$.

Thus the smallness of the neutrino masses and the stringent
proton decay lifetimes demand that $x_i, y_i \ll 1$. It is
impossible to accomplish this from Eqns. (\ref{Mi}) and
(\ref{mui}) by any choice of $\eta_i$. So extension of bilinear
RPV to SUSY SU(5) has a serious difficulty. 

One way to bypass this conundrum is to keep the trilinear RPV
terms, $Y^5_{ijk}$, in Eqn. (\ref{master}) to be non-zero 
and make additional fine-tunings so that they almost exactly
cancel off a large contribution emerging from the $x_i$ leaving a
small remainder consistent with proton decay and further ensure
through other fine-tunings that  the neutrino
mass remains small enough \cite{Bajc:2015zja}.  Alternatively, one
may abandon the principle of naturalness and set $\eta_i = 0$ so
that ${\cal M}_i = \mu_i = M_i$ and choose $M_i$ satisfying the
bound on $y_i$ from the neutrino mass\footnote{$\eta_0 \neq
0$ produces doublet-triplet splitting, i.e.,  ${\cal M}_0 \gg
\mu_0$. So, now $M_i/{\cal M}_0 = x_i \ll y_i = M_i/\mu_0$.}.
The limits on $x_i$  then imply that the matrix $U$
like $V$ is almost an unit matrix.  Hence, $m^{\rm diag} \approx
m_d = diag(m_{d_i})$. However, $m^{\rm diag}
\approx diag(m_{\ell_i})$ from discussions following
Eqn.~(\ref{subchargino}). So finally, $m_{d_i}\approx
m_{\ell_i}$, or in other words, bilinear RPV couplings ${\cal
M}_i$ and $\mu_i$ do not help correct the wrong fermion mass
ratio problem posed by RPC SUSY SU(5) scenario.

\section{RPV cannot solve the colour triplet mass problem}
As MSSM is promoted to SUSY SU(5), RG running of the gauge coupling constants are affected by the mass of the colour triplets $T$ and
$\overline{T}$.  
Thus, the requirement of gauge coupling unification relates  the GUT
scale with $M_T$. We can roughly see at 90\% CL~\cite{Murayama:2001ur},
\beq
3.5\times 10^{14} < M_T ~{\rm (GeV)} < 3.6 \times 10^{15} \;,
\eeq
whereas the proton decay constraints puts a lower bound on $M_T$:
\beq
M_T > 7.6 \times 10^{16} ~{\rm GeV} \;.
\eeq
It leads to a discrepancy in the minimal SUSY SU(5) framework. It
is interesting to explore whether RPV can at least provide a solution to
this problem.

In ~\cite{Allanach:1999mh} the changes in the grand
unification scale due to the presence of RPV couplings have been
explored. It has been shown for order one RPV couplings at the
electroweak scale the change in $M_{GUT}$ is at the most
20\%. Hence, it cannot solve the disparity in the above two
bounds, unless the RPV contribution to proton decay destructively
interferes  in a fine-tuned manner with colour triplet
mediated proton decay.

In the situation discussed here, namely, that the sole RPV is
generated from the ${\cal M}_i$ terms, the $\lambda^\prime$ and
$\lambda^{\prime\prime}$ couplings so produced are 
constrained from proton decay to be way too small to make any
appreciable effect on gauge coupling unification to address the
above disparity.

\section{Summary and Conclusion}
\label{S:conclusion}

Promoting $R$-parity violation from the  MSSM to SUSY SU(5)
introduces a new type of bilinear RPV coupling ${\cal M}_i$
involving the down-type antiquarks of the matter $\bar{5}$-plets
and the superpartner of SU(2) singlet, colour anti-triplets $\overline{T}$ which  are
members of the SU(5) Higgs $\bar{5}$-plet.  This is in addition
to the usual bilinear RPV terms, $\mu_i$, that induce mixing
between the SU(2) doublet leptons and the Higgsino.

The mixing among the colour anti-triplet states resulting from
the diagonalisation of the mass matrix introduces trilinear RPV
$\lambda^\prime$ and $\lambda^{\prime\prime}$ terms which can
lead to proton decay. The strong bounds ensuing from
non-observation of proton decay in experiments such as
SuperKamiokande restrict these RPV couplings
to  ${\cal M}_i/{\cal M}_0 \sim 10^{-7}$. 
On the other
hand  the $\mu_i$ terms result in neutrino-neutralino mixing at
the tree level. The smallness of the neutrino masses {\em
vis-\`{a}-vis} the weak scale implies that the ratio
$\mu_i/\mu_0$ are also quite suppressed -- $\sim 10^{-6}$.

The type of fine-tuning that introduces doublet-triplet mass
splitting within the SU(5) Higgs $\bar{5}$-plet can be extended
to the RPV sector. This can suppress either the
$\overline{T}-d^c$ mixing or the standard bilinear RPV couplings
involving leptons, but not both simultaneously.  This is an
obstacle to extending RPV to SUSY SU(5).

One way to circumvent this impasse is to assume  further
fine-tuned cancellations across different sectors by (a)
introducing trilinear RPV terms in the Lagrangian which precisely
compensate the ones generated through the mechanism above to
leave a small remnant that is consistent with proton decay
limits, and (b) ensure through a different set of fine-tunings
that the tree- and loop-level contributions to the neutrino mass
remain under control.

One may instead not abide by the principle of naturalness, take
$\eta_i = 0$, and  choose $M_i$ so that $y_i = {M}_i/\mu_0 \sim
10^{-6}$. Then $x_i = M_i/{\cal M}_0 \sim 10^{-20}$ is utterly
negligible.   SU(5) symmetry dictates $m_{d_i}=m_{\ell_i}$ as both
down-type quark and charged lepton Yukawa couplings originate
from the $Y^5$ term in the SUSY SU(5) superpotential in
Eqn.~(\ref{master}).  Due to the presence of ${\cal M}_i$ and
$\mu_i$, it may appear that the mass ratios $m_{d_i}/m_{\ell_i}$
can be altered as desired by adjusting these terms.  However,
because of the tight constraint on $x_i$  the
change in the down-type quark masses are not appreciable.  The
weak scale masses in $m^{\rm diag}$ in Eqn.~(\ref{quarkT}) also
induce a small mixing $\sim {\cal O}(M_W/M_{GUT})$ between the
left-handed down-type quarks with $T$. The effect of such small
mixings on the down-type quark masses is insignificant.  The
usual bilinear RPV terms, $\mu_i$, tightly constrained by the
size of the neutrino mass, induce mixing between the charged
leptons and the charged Higgsino. This also leads to a
modification in the charged lepton masses but it is negligibly
small.

In summary,  promoting bilinear RPV to SUSY SU(5) faces a
naturalness obstacle from the twin limits on proton decay
lifetime and the neutrino mass. Extending the fine-tuning that
ensures doublet-triplet splitting in the Higgs multiplet to SUSY
 does not provide a solution. One way out is to abandon the
naturalness principle itself. An alternate possibility is to
invoke  several further fine-tunings in sectors which are
{\em a priori} unrelated. Without such a procedure one  cannot
change the SU(5) prediction of the mass ratio
$m_{d_i}/m_{\ell_i}$ significantly.

\section{Acknowledgements}
N.K. acknowledges financial support from University Grants
Commission, India.  A.R. is partially funded by  the Department of
Science and Technology Grant No. SR/S2/JCB-14/2009.

\appendix 
\section{The mixing matrix}
\label{appendix1}
In case only one generation of fermions is considered,
Eqn.~(\ref{quarkT}) looks like
\beq
\begin{pmatrix}
\overline{T}_0 & d^c_0
\end{pmatrix}
\begin{pmatrix}
{\cal M}_0 & 0 \\ 
{\cal M}_1 & m_1\\ 
\end{pmatrix}
\begin{pmatrix}
{T}_0\\ d_0 
\end{pmatrix} 
\equiv 
\begin{pmatrix}
\overline{T}_0 & d^c_0
\end{pmatrix}
\mathscr{M}
\begin{pmatrix}
{T}_0\\ d_0 
\end{pmatrix} \; ,
\label{flavour1a}
\eeq
where $m_1=Y^5_{1}\,v_d$.

The mass matrix $\mathscr{M}$ can be diagonalised by a bi-unitary
transformation:
\begin{equation}
\begin{pmatrix}
\overline{T}&& d^{c} 
\end{pmatrix}
\begin{pmatrix}
 M_T & 0 \\ 
0 & m_d\\ 
\end{pmatrix}
\begin{pmatrix}
T\\ d
\end{pmatrix} =  \begin{pmatrix}
\overline{T}&& d^{c} 
\end{pmatrix}
\left[U_R \mathscr{M} U_L^\dagger \right] \begin{pmatrix}
T\\ d
\end{pmatrix} \;,
\label{biunit}
\end{equation}
where $U_R \mathscr{M} \mathscr{M}^\dagger U_R^\dagger =
U_L \mathscr{M}^\dagger \mathscr{M} U_L^\dagger = 
diag\{M_T^2 ~~m_d^2\}$ and the mass basis eigenstates are: 
\begin{eqnarray}
\begin{pmatrix}
\overline{T}\\ d^{c} 
\end{pmatrix} &=& U_R \begin{pmatrix}
\overline{T}_0\\ d^{c}_0 
\end{pmatrix} = 
\begin{pmatrix}
c_R & s_R \\ 
-s_R & c_R\\ 
\end{pmatrix}
\begin{pmatrix}
\overline{T}_0\\ d^{c}_0 
\end{pmatrix} \;,
\nonumber \\
\begin{pmatrix}
T \\ d 
\end{pmatrix} &=& U_L \begin{pmatrix}
T_0\\ d_0 
\end{pmatrix} = 
\begin{pmatrix}
c_L & s_L \\ 
-s_L & c_L\\ 
\end{pmatrix}
\begin{pmatrix}
T_0\\ d_0 
\end{pmatrix} \;.
\label{transform1}
\end{eqnarray}
Above,  $c_{L,R} = \cos\theta_{L,R}$, $s_{L,R}=\sin\theta_{L,R}$. 
The mass eigenvalues (ignoring terms of ${\mathcal O}(m_1^4)$) are:
\begin{equation}
M_T = {\cal M}_0 \sqrt{1 + x^2 + \frac{x^2 z^2}{1+x^2}} \simeq {\cal M}_0
\sqrt{1 + x^2}, \;\; 
m_d = {\cal M}_0 \sqrt{\frac{z^2}{1+x^2}} \simeq 0 \;\; ,
\end{equation}
where $x = {\cal M}_1/{\cal M}_0$ and $z = m_1/{\cal M}_0 \ll x$ 
since $m_1\sim{\cal O}(M_W)$ and ${\cal M}_0\sim {\cal M}_1\sim
{\cal O}(M_{GUT})$. The mixing angles are given by:
\beq
\tan{2\theta_R} = \frac{2x}{1 - x^2 - z^2} \simeq \frac{2x}{1-x^2}
\qquad {\rm and } \qquad 
\tan{2\theta_L} = \frac{2 zx}{1 + x^2 - z^2} \simeq 0 \;.
\label{thetaa}
\eeq
Thus, the mixing between $T$ and $d$ is
negligible while that between $\overline{T}$ and $d^c$ can be
significant. Indeed, as $x \rightarrow 1$ the  mixing angle
$\theta_R$ tends to its maximal value of $\pi/4$. On the other
hand for the $x \rightarrow 0$ limit, as expected $\theta_R \rightarrow 0$. 

From Eqn. (\ref{thetaa}) to a good approximation:
\begin{equation}
\cos \theta_R = \frac{1}{\sqrt{1 + x^2}}, \;\; \sin \theta_R =
\frac{x}{\sqrt{1 + x^2}}, \;\;
\cos \theta_L = 1, \;\; \sin \theta_L = 0    \;\;.
\label{mix1g}
\end{equation}
This result can be readily extended to three generations.



\vskip 20pt
  
\begin{thebibliography}{99}

\bibitem{RPV-neutrino} 
  L.~J.~Hall and M.~Suzuki,
  Nucl.\ Phys.\ B {\bf 231}, 419 (1984).
For a review see, for example, 
  O.~C.~W.~Kong,
  Int.\ J.\ Mod.\ Phys.\ A {\bf 19}, 1863 (2004)
  [hep-ph/0205205]. 
 
\bibitem{Rakshit:2004rj} 
  S.~Rakshit,
  Mod.\ Phys.\ Lett.\ A {\bf 19}, 2239 (2004)
  [hep-ph/0406168].
 
\bibitem{Bajc:2015zja} 
  B.~Bajc and L.~Di Luzio,
  JHEP {\bf 1507}, 123 (2015)
  [arXiv:1502.07968 [hep-ph]].

\bibitem{DiazCruz:2000mn} 
  J.~L.~Diaz-Cruz, H.~Murayama and A.~Pierce,
  Phys.\ Rev.\ D {\bf 65}, 075011 (2002)
  [hep-ph/0012275].

\bibitem{Babu:2012pb} 
  K.~S.~Babu, B.~Bajc and Z.~Tavartkiladze,
  Phys.\ Rev.\ D {\bf 86}, 075005 (2012)
  [arXiv:1207.6388 [hep-ph]].
  
\bibitem{others} 
  J.~R.~Ellis and M.~K.~Gaillard,
  Phys.\ Lett.\ B {\bf 88}, 315 (1979);
  H.~Georgi and C.~Jarlskog,
  Phys.\ Lett.\ B {\bf 86}, 297 (1979).

\bibitem{Nishino:2009aa} 
  H.~Nishino {\it et al.}  [Super-Kamiokande Collaboration],
  Phys.\ Rev.\ Lett.\  {\bf 102}, 141801 (2009)
  [arXiv:0903.0676 [hep-ex]].

\bibitem{Abe:2014mwa} 
  K.~Abe {\it et al.}  [Super-Kamiokande Collaboration],
  Phys.\ Rev.\ D {\bf 90}, 072005 (2014)
  [arXiv:1408.1195 [hep-ex]].

\bibitem{protondecay} 
  A.~Y.~Smirnov and F.~Vissani,
  Phys.\ Lett.\ B {\bf 380}, 317 (1996)
  [hep-ph/9601387];

  C.~E.~Carlson, P.~Roy and M.~Sher,
  Phys.\ Lett.\ B {\bf 357}, 99 (1995)
  [hep-ph/9506328];

  I.~Hinchliffe and T.~Kaeding,
  Phys.\ Rev.\ D {\bf 47}, 279 (1993).
  
\bibitem{Weinberg:1981wj} 
  S.~Weinberg,
  Phys.\ Rev.\ D {\bf 26}, 287 (1982).

\bibitem{Sakai:1981pk} 
  N.~Sakai and T.~Yanagida,
  Nucl.\ Phys.\ B {\bf 197}, 533 (1982).

\bibitem{Bajc:2002bv} 
  B.~Bajc, P.~Fileviez Perez and G.~Senjanovic,
  Phys.\ Rev.\ D {\bf 66}, 075005 (2002)
  [hep-ph/0204311].

\bibitem{Smirnov:1995ey} 
  A.~Y.~Smirnov and F.~Vissani,
  Nucl.\ Phys.\ B {\bf 460}, 37 (1996)
  [hep-ph/9506416].

\bibitem{Bhattacharyya:1997vv} 
  G.~Bhattacharyya,
  In *Tegernsee 1997, Beyond the desert 1997* 194-201
  [hep-ph/9709395].

\bibitem{Bhattacharyya:1996nj} 
  G.~Bhattacharyya,
  Nucl.\ Phys.\ Proc.\ Suppl.\  {\bf 52A}, 83 (1997)
  [hep-ph/9608415].


\bibitem{Barbier:2004ez} 
  R.~Barbier {\it et al.},
   Phys.\ Rept.\  {\bf 420}, 1 (2005)
   [hep-ph/0406039];

   see also   I.~Hinchliffe and T.~Kaeding, in Ref. \cite{protondecay}. 
  
 
\bibitem{Murayama:2001ur} 
  H.~Murayama and A.~Pierce,
  Phys.\ Rev.\ D {\bf 65}, 055009 (2002)
  [hep-ph/0108104].
  
\bibitem{Allanach:1999mh} 
  B.~C.~Allanach, A.~Dedes and H.~K.~Dreiner,
  Phys.\ Rev.\ D {\bf 60}, 056002 (1999)
  [Phys.\ Rev.\ D {\bf 86}, 039906 (2012)]
  [hep-ph/9902251].

   
\end{thebibliography}
\end{document}